\documentclass[osajnl,twocolumn,showpacs,superscriptaddress,10pt]{revtex4-1} 
\usepackage{amsmath,amssymb,graphicx}
\begin{document}

\title{Tunable Geometric Fano Resonances in a Metal/Insulator Stack}

\author{Herbert Grotewohl}
\author{Miriam Deutsch}\email{Corresponding author: miriamd@uoregon.edu}
\affiliation{Department of Physics, University of Oregon, 1371 East 13th Street, Eugene, Oregon 97403, USA}

\begin{abstract}
\noindent We present a theoretical analysis of surface-plasmon-mediated mode-coupling in a planar thin film metal/insulator stack. Solving for the modes of this structure, we obtain an avoided crossing in the angular domain that is tunable using the system's materials and geometrical parameters. The computed reflectance of the metal/insulator stack exhibits a lineshape that is modeled as a geometric Fano resonance, accompanied by a response function characteristic of an induced coherence, albeit in the angular domain. We also observe a reversal of the Fano lineshape asymmetry as the system's parameters are varied. We attribute this \emph{q}-reversal to an interloping background radiation field comprising the surface plasmon polariton mode.
\end{abstract}


\maketitle 

The interference of a discrete state with a broad background is known to give rise to Fano resonances. Originally, Fano resonances were observed in the autoionization spectrum of helium, explained as outcome of quantum interference between scattering amplitudes of a bound state and a continuum~\cite{Fano1961}. The rapid variation in phase and amplitude, characteristic of a resonant state, gives rise to the well-known asymmetric Fano lineshape. While Fano resonances were initially described as quantum interference effects in atomic systems, their presence has since been demonstrated in coupled nanoscale structures~\cite{Kobayashi2004,Miroshnichenko2010}, plasmonic systems~\cite{Halas2010,Lu2012,Lovera2013}, as well as coupled classical oscillators~\cite{Satpathy2012}. Studies to date have predominately addressed Fano interference in the frequency domain, often utilizing engineered metamaterials to control the frequency response~\cite{Giessen2009,Luk'yanchuk2010}. Alternately, manipulation of quantum interference in \emph{space} using geometrical parameters has been demonstrated to enable control of resonant optical mode coupling in guided-wave structures~\cite{Herminghaus1994,Rotenberg2013} as well as nano-scale plasmonic systems~\cite{Rohde2007}, in manners analogous to Fano interference and electromagnetically induced transparency (EIT)~\cite{Marangos1998}. Such structural control demonstrates quantum interference of \emph{geometrical resonances}, thus enabling the extension of many known aspects of quantum coherence theory to the spatial domain.

In this work we analyze a geometric Fano resonance, arising from interference between a Surface Plasmon Polariton (SPP) mode and a metal-clad dielectric waveguide (metal-insulator-metal or MIM) mode. 
The system consists of two planar and infinite metal films of permittivity $\epsilon_m$ sandwiching a thin insulator with permittivity $\epsilon_g$, thus forming a metal/insulator stack (MIS). The remaining two metal interfaces of the MIS each abut a semi-infinite, isotropic and homogeneous dielectric host with permittivity values of $\epsilon_p$ and $\epsilon_e$, respectively. The SPP mode considered here is excited at the metal/$\epsilon_e$ interface. The MIM mode is a leaky guided mode, confined to the insulating film~\cite{Shin2004,Dionne2006a}.

More specifically, we address a geometric Fano resonance (GFR) of the system in the optical frequency range, utilizing MIS structures where all film thicknesses range mostly between 50-100nm. We solve for the modes of the system and demonstrate their spectral tuning through geometrical and materials parameters. We obtain an angular response function, analogous to the susceptibility of a dressed atom exhibiting EIT, that indicates the presence of a geometric coherence underlying the GFR. We then calculate the complex Fresnel reflection coefficients, and show that using the aforementioned parameters it is possible to tune both the angular lineshape as well as the resonance frequency of the GFR.

To analyze the GFRs of this system, we begin with computing the modes of a simpler structure where the dielectric with permittivity $\epsilon_p$ is replaced with a semi-infinite metal film, as shown schematically in the Inset to Fig. 1. This approach isolates the MIM and SPP fields from most radiation modes, thus allowing us to first analyze mode coupling effects in the unperturbed system. The dispersion relations are obtained from Maxwell's equations by applying the appropriate boundary conditions at the planar interfaces. The solutions are thus presented as coupled modes of the two subsystems comprising the structure, namely a MIM with two semi-infinite metal films (labeled \emph{infinite MIM}) and a single metal/dielectric interface. Coupling is obtained by bringing these two subsystems into close proximity, such that a metal film of thickness $t_c$ enables mode overlap~\cite{Rohde2007}. It is important to note here that the eigenvalues associated with metal-supported modes are in general complex, a consequence of material loss and the radiative nature of MIM modes~\cite{Shin2004}. Without loss of generality, we set $k_{||}$, the in-plane wave vector to be real-valued, thus rendering the eigenfrequencies complex~\cite{Sernelius}.

Using this approach, we obtain the transverse magnetic (TM) eigenmodes of the isolated system. (While the MIM structure also supports transverse electric solutions, the latter do not couple to SPP modes, which are solely TM in nature.) For both metal films we choose a silverlike Drude metal with dispersion $\epsilon_m(\omega)\equiv \epsilon_b - \omega_p^2 (\omega^2 + i \Gamma \omega)^{-1}$~\cite{Gadenne1998}. Here $\epsilon_b = 5.1$; $\hbar \omega_p = 9.1$ eV, with $\omega_p$ the bulk plasma frequency, and $\hbar \Gamma = 0.021$ eV. The external insulator is chosen as vacuum ($\epsilon_e=1$). For the waveguide insulator film we use titania in rutile form, with a dispersive permittivity  $\epsilon_g = \epsilon_{TiO_2}(\omega)$\cite{DEVORE1951}. 
The thickness $t_c$ of the coupling metal-film determines the spatial overlap of the SPP and MIM modes, and may be varied to control the interaction between these two fields.

\begin{figure}[t]
\begin{minipage}{1\columnwidth}
  \leftline{\includegraphics[width=\columnwidth]{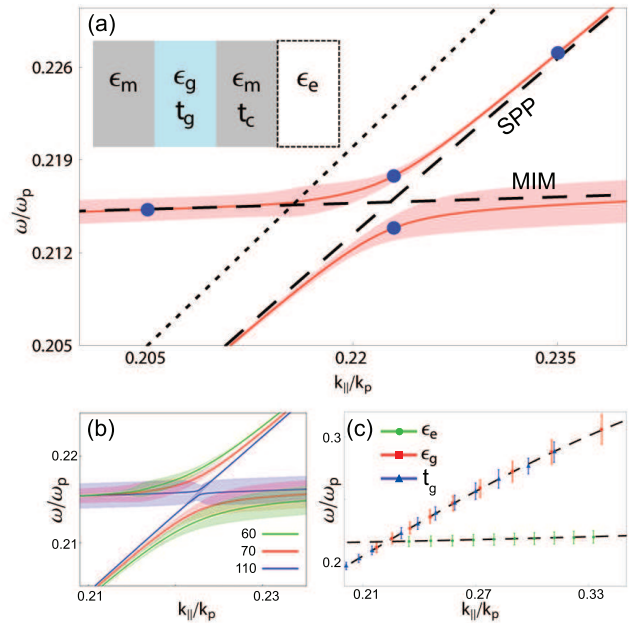}}
\end{minipage}
\caption{(a) Mode dispersions showing the SPP and infinite MIM modes (dashed), coupled MIS modes (solid broad traces), and vacuum light line (dotted). Here $k_p\equiv\omega_p/c$ and $k_{||}$ is the in-plane wavenumber. Inset: Schematic of the structure used to compute the plotted modes, showing an infinite MIM coupled to an external dielectric with permittivity $\epsilon_e$ through a metal film with thickness $t_c$. The metal film on the left-hand of the structure is infinite in thickness. (b) Mode dispersions for metal coupling films of thicknesses as denoted (in nm), using $t_g=80$nm and $\epsilon_e=1$. (c) Location of avoided crossing (points) and minimal frequency separation (bars), computed using the following parameter ranges: $1<\epsilon_e<2$, $2.5<\epsilon_g<7$, and $40<t_g<100$.  The infinite MIM and SPP modes computed using parameter values as in (a) are shown as dashed lines.}
\label{fig:modes}
\end{figure}

In Fig.~\ref{fig:modes}(a) we plot the first order coupled frequency eignemodes of the system using $t_c=70$nm. The dashed traces (labeled) represent the uncoupled modes of the two subsystems, as described above. The dotted line indicates the vacuum light line. As demonstrated previously~\cite{Hasegawa2006}, the geometric resonance condition $t_g\approx\lambda_{sp}/4\sqrt{\epsilon_g}$, where $\lambda_{sp}$ is the free-space surface plasmon wavelength, may be used to obtain the optimal insulator thickness resulting in a nearly-flat MIM mode. This condition yields a broad angular resonance that enables omnidirectional coupling to MIM modes~\cite{Shin2004}.  Application of the geometric resonance condition gives $t_g=44$nm, which renders the crossing with the SPP mode close to the ultraviolet region of the spectrum. In order to facilitate a more practical interrogation wavelength, we increase the insulator thickness to $t_g=80$nm. This red shifts the intersection point to near $\lambda=632.8$nm ($\omega/\omega_p = 0.216$), while maintaining a nearly flat MIM mode over the entire angular range of interest.  The effects of varying $t_g$ are discussed further below.


The broad traces in Fig.~\ref{fig:modes}(a) show the modes of the coupled MIS, plotted using $t_g=80$nm and $t_c=70$nm. The solid center line in each trace denotes the real value of the complex eigenfrequency, while the width of each resonance is given by the shaded region, equal in magnitude to twice the imaginary part of the frequency eigenvalue. The observed level repulsion at $\omega/\omega_p\approx0.216$ is a result of the coupling between the infinite MIM mode and the SPR at the single metal/insulator interface, enabled now through the finite value of $t_c$. Plotting the real and imaginary values for the eigenfrequencies allows us to better analyze the effect of geometrical parameters as well as losses on mode coupling, and subsequently on the GFR. This is discussed in more detail below.

The coupling between the modes is controlled via tunable geometrical and materials parameters, namely the thicknesses of both metal and insulator films, as well as the permittivities of both waveguide and external insulator materials. In Fig.~\ref{fig:modes}(b) we plot the real and imaginary values of the eigenmodes of the coupled MIS for several thicknesses of the coupling metal film, using a fixed value of $t_g=80$nm. As $t_c$ increases, center-to-center mode separation decreases and the traces tend more closely towards the uncoupled modes. This is due to the diminished overlap experienced by the fields when screened by an increasingly thick metal film, resulting in weaker hybridization. We note here that we do not investigate the effect of metal films significantly thinner than 60nm, since at lower values of $t_c$ field overlap between the interfaces is very high, and the modes do not retain relevant attributes of the uncoupled solutions. Examining the widths of the modes, we find that generally SPR-like modes are narrower than the MIM-like solutions. This is because while the SPR is a lossy evanescent mode that is bound to a single metal/insulator interface, the MIM mode is tightly confined between two such interfaces, thus experiencing shorter propagation lengths and increased losses at the insulator thickness chosen here~\cite{Zia2004}. We also find that at increasingly large values of the wavevector all MIM-like modes exhibit relatively large and comparable widths, indicating significant losses. This is expected, as greater values of $k_{||}/k_p$ indicate increasing interaction of the propagating field with the metal layers, resulting in higher extinction~\cite{Preiner2008}. 

In addition to the decreased center-to-center separation between the modes, increasing $t_c$ also results in overlap of mode widths in the coupling region, as seen in Fig.~\ref{fig:modes}(b). Effectively, greater values of $t_c$ lead to increased attenuation of the propagating coupled fields, accompanied by strong screening of the two modes~\cite{Zia2004}. Close to the coupling region, we observe a change in the width, as the modes evolve from MIM-like to SPR-like, and vise versa. The stronger the hybridization of the modes the more gradual the change in width. For the thickest films, where the modes are strongly screened, this change in width happens abruptly, very close to the coupling region, as expected.

\begin{figure}[t]
\begin{minipage}{1\columnwidth}
  \leftline{\includegraphics[width=\columnwidth]{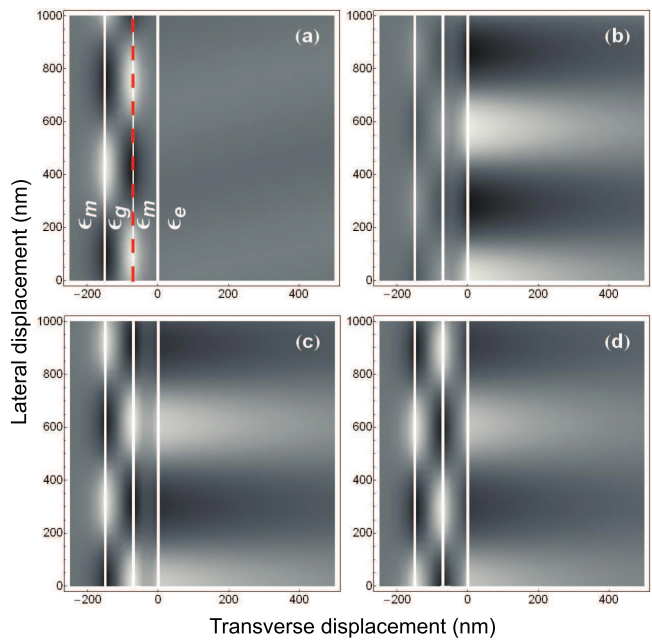}}
\end{minipage}
\caption{Real part of the magnetic field of the modes, plotted for values of $k_{||}/k_p$ equal to (a) 0.205, (b) 0.235, (c) 0.223 (antisymmetric mode, $\omega/\omega_p\approx0.218$), (d) 0.223 (symmetric mode, $\omega/\omega_p\approx0.214$). The composition of each layer is denoted in (a) and is identical in all panels. The dashed line in (a) designates the reference plane with respect to which the symmetry of the modes is determined.}
\label{fig:hfields}
\end{figure}

While $t_c$, the thickness of the coupling metal film serves as the primary control over the strength of the interaction, the dispersion of the coupled system may be tuned through the remaining geometrical parameters comprising the thickness, $t_g$ and permittivity, $\epsilon_g$ of the insulating film, as well as the permittivity $\epsilon_e$ of the external insulator. In Fig.~\ref{fig:modes}(c) we plot the position of the avoided crossing, determined as the midpoint value of the minimal frequency separation between the two branches of the coupled MIS modes. The minimal difference between the frequency values is also used to designate the height of the bar for each frequency value plotted. (Only the real values of the eigenfrequencies are used to obtain these plots.) We find that variations to the insulating film's parameters yield very similar results, indicating that the main effect of the insulating film on the GFR is to alter the relative phase of the coupled modes through its optical thickness. As already mentioned above, increasing the optical thickness of the insulator redshifts the intersection of the MIM and SPR modes, resulting in the observed monotonic decrease shown in Fig.~\ref{fig:modes}(c). In addition, increased confinement of the MIM mode in higher optical thickness films results in weaker hybridization with the SPR mode, and hence smaller gap values. Alternately, the main effect of increasing $\epsilon_e$ is to shift the SPR mode further to the right of the light line, without much impact on the MIM resonance (for the value of $t_c$ used here.) Thus we find the GFR following this shift to where the SPR and MIM now intersect at higher values of $k_{||}$, as also seen in Fig.~\ref{fig:modes}(c). This shift does not impact the hybridization strength, as demonstrated by the constant magnitude of the frequency gap. 

The evolution of the modes from MIM-like to SPR-like is demonstrated by plotting the magnetic field distributions, as shown in Fig.~\ref{fig:hfields}. Each plot corresponds to one of the four dots marking the modes in Fig.~\ref{fig:modes}(a).
Panels (a) and (b) of Fig.~\ref{fig:hfields} shows predominate excitation of the MIM and SPP modes, respectively, as expected when the modes are far detuned from the GFR. On resonance, when $k_{||}/k_p=0.223$, mode coupling results in a pair of eigenmodes of opposite symmetries, as seen in panels (c) and (d) of Fig.~\ref{fig:hfields}.

\begin{figure}[t]
\begin{minipage}{\columnwidth}
  \centerline{\includegraphics[width=\columnwidth]{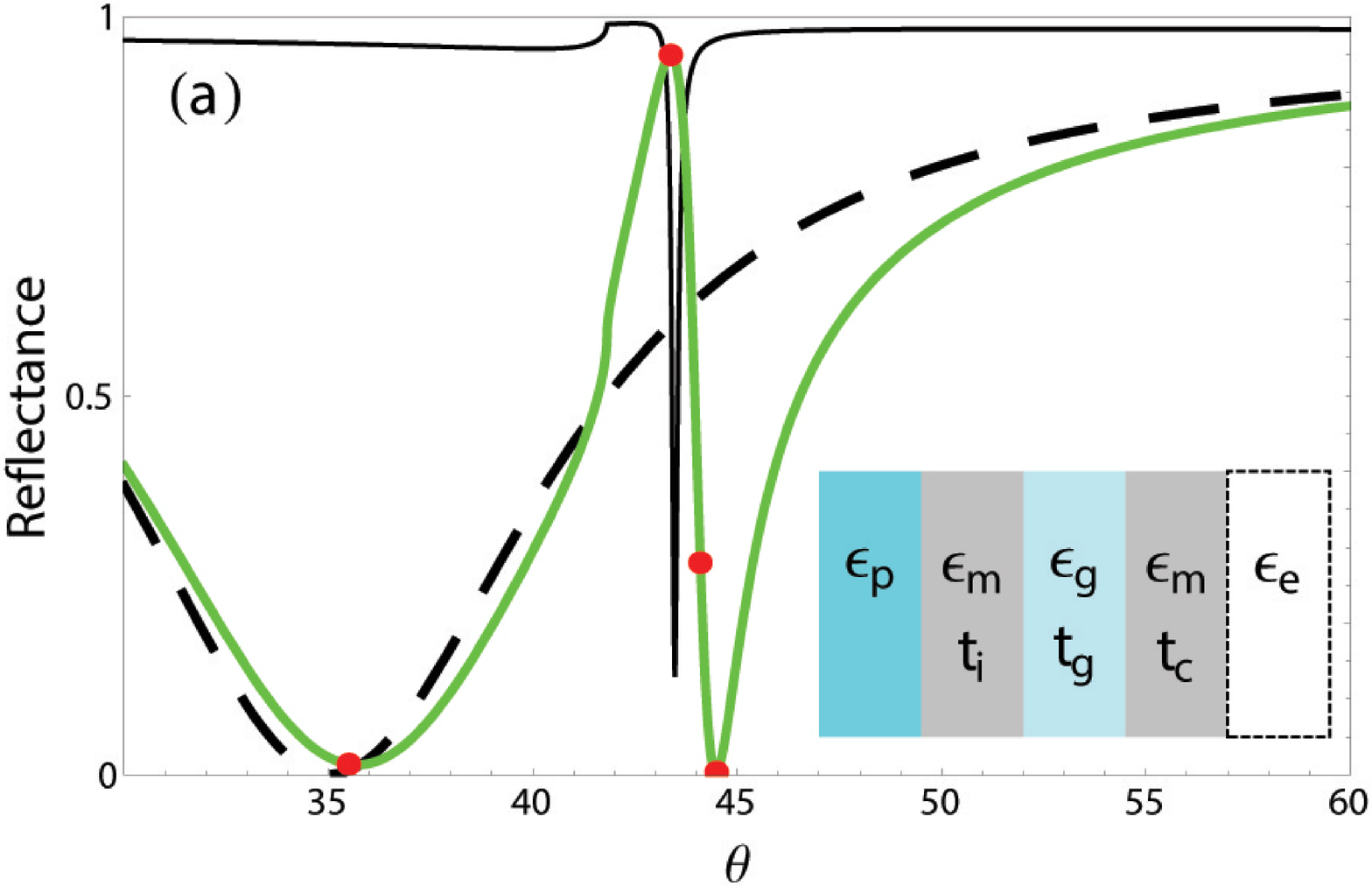}}
\end{minipage}
\begin{minipage}{\columnwidth}
  \centerline{\includegraphics[width=\columnwidth]{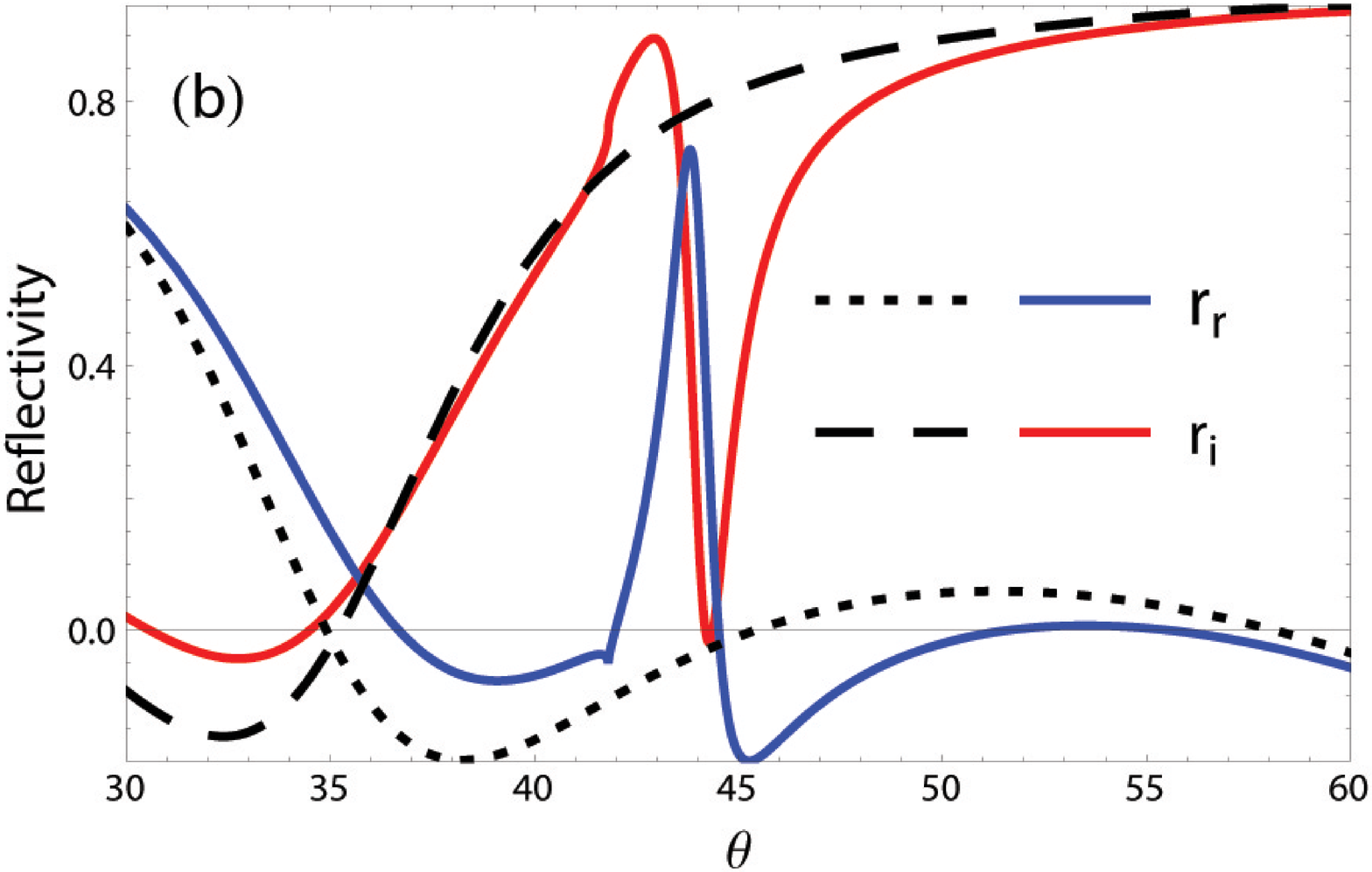}}
\end{minipage}
\caption{(a) Computed reflectance of the MIS with $t_i=55$nm, $t_g=80$nm and $t_c=70$nm, exhibiting a GFR (heavy solid trace). Also shown are reflectances of the semi-infinite MIM (dashed trace) and a metal film of thickness $t_i=55$nm (thin solid trace). Here $\epsilon_p=2.25$, $\epsilon_g=\epsilon_{TiO_2}$ and $\epsilon_e=1$. Inset: Schematic of the MIS structure. The medium with permittivity $\epsilon_p$ may be shaped as a prism, to facilitate coupling into the SPR mode. (b) Reflected field quadratures, $r=r_r+i r_i$, of the GFR (solid traces) and of the semi-infinite MIM (dashed and dotted traces).}
\label{fig:Randr}
\end{figure}

We now return to the complete MIS structure comprising the semi-infinite dielectric layer of permittivity $\epsilon_p$, as shown schematically in the inset to Fig.~\ref{fig:Randr}(a). The semi-infinite metal film is now rendered a film of finite thickness $t_i$. In practical configurations, the dielectric $\epsilon_p$ is often structured as a prism, resulting in the Kretschmann excitation scheme for SPPs~\cite{Herminghaus1994}. We note that the $\epsilon_p$/metal interface supports an additional SPP mode, which also couples to the MIM mode. However, this interaction occurs at values of $k_{||}/k_p$ lying outside the lightcone as defined now by $\epsilon_p$, and is therefore not accessible using the current scheme.

In Fig.~\ref{fig:Randr}(a) we plot the reflected intensity of a plane wave impinging from $\epsilon_p$ on the structure shown in its Inset, as a function of the incidence angle $\theta$, for a light field with $\omega/\omega_p=0.216$. Values of films parameters are the same as those used to compute the modes in Fig.~\ref{fig:modes}(a), with the addition of $t_i=55$nm. For the coupling dielectric we use $\epsilon_p= 2.25$, the permittivity of fused silica at the chosen frequency. In analogy to the subsystems used above to compute the modes, we also compute the reflectances of a semi-infinite MIM (formed by coupling the infinite MIM to $\epsilon_p$ through $t_i$) and a metal film of thickness $t_i=55$nm sandwiched between $\epsilon_p$ and vacuum. All incidences are from $\epsilon_p$, for plane wave fields at $\omega/\omega_p=0.216$. These computed reflectances are shown for reference in Fig.~\ref{fig:Randr}(a) as dashed and thin solid traces, respectively. The sharp dip in the single-film reflectance indicates excitation of the SPR at the metal/vacuum interface. The critical angle for total internal reflection is $\theta_{cr}=41.8^\circ$, as evidenced by the cusp in the SPR trace, as well as in the GFR. The reflectance of the entire MIS structure exhibits the asymmetric lineshape typical of Fano resonance, albeit here it is manifest in the angular domain. Unlike many common Fano systems, where the response approaches a constant value when the system is far detuned from resonance, here we find that the reflectance follows the MIM trace at angles far from the SPR resonance. This is to be expected, as the non-resonant response is determined by the dispersion of the broad continuum. While the latter is often nearly-flat in atomic systems, in our structure it is the dispersive MIM mode which plays the role of the background, thus determining the overall angular dispersion of the MIS.

To verify the coherent nature of the GFR we also compute the quadratures of the reflected light field. The latter are plotted in Fig.~\ref{fig:Randr}(b), along with the reflectance quadratures for the semi-infinite MIM defined above. We find that the real and imaginary components of the GFR field indeed follow a response function typical of a dressed system~\cite{Xiao1995}, indicating the existence of coherent angular response. This is highlighted by the dashed traces representing the MIM quadratures, where we see that the angular coherence results in modification of the background response only in vicinity of the SPR, as expected.

\begin{figure}[t]
\begin{minipage}{\columnwidth}
  \centerline{\includegraphics[width=\columnwidth]{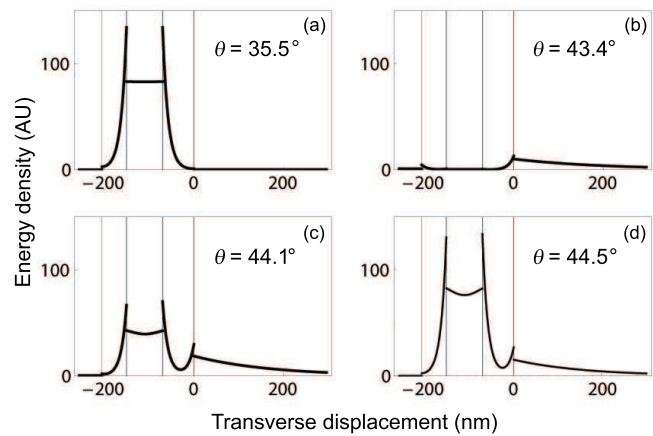}}
\end{minipage}
\caption{Cross-sectional time averaged EM energy density, computed relative to the incident energy density. The origin of all abscissae is at the $\epsilon_m/\epsilon_e$ interface. Incidence angles at which the energy densities were computed are also stated.}
\label{fig:edensity}
\end{figure}

The reflectance lineshape of the GFR can be further understood by examining the EM energy density distributions in the MIS. To obtain these, we compute the EM energy density in each film separately while accounting appropriately for the losses and dispersion in the metal~\cite{LandauandLifshitz}. In Fig.~\ref{fig:edensity} we show the cross-sectional time averaged EM energy density for incidence angles designated by the four points marking the GFR trace in Fig.~\ref{fig:Randr}(a). At $\theta=35.5^\circ$ we find that all the energy in the MIS resides in the MIM mode, as suggested by the vanishing reflectance at this angle. (Coupling into the SPR is not yet possible since this angle is lower than the critical angle for total internal reflection.) Examining the energy densities at greater angles, we find that above $\theta_{cr}$ energy couples into both the MIM and SPR modes, at ratios determined by their relative phase. In particular, at $\theta=43.4^\circ$ where the reflectance is maximal we observe the SPR being predominately excited, with negligible energy in the MIM mode. This implies coherent suppression of the MIM mode, in analogy to EIT, as well as to dynamic damping observed in other classical coupled oscillator systems~\cite{Tokman2002}. As the angle increases further, the relative phase between the interfering MIM and SPR modes changes with the increasing optical path, resulting in various degrees of excitation of the two modes.

\begin{figure*}[t]
\begin{minipage}{.3\textwidth}
  \centerline{\includegraphics[width=\columnwidth]{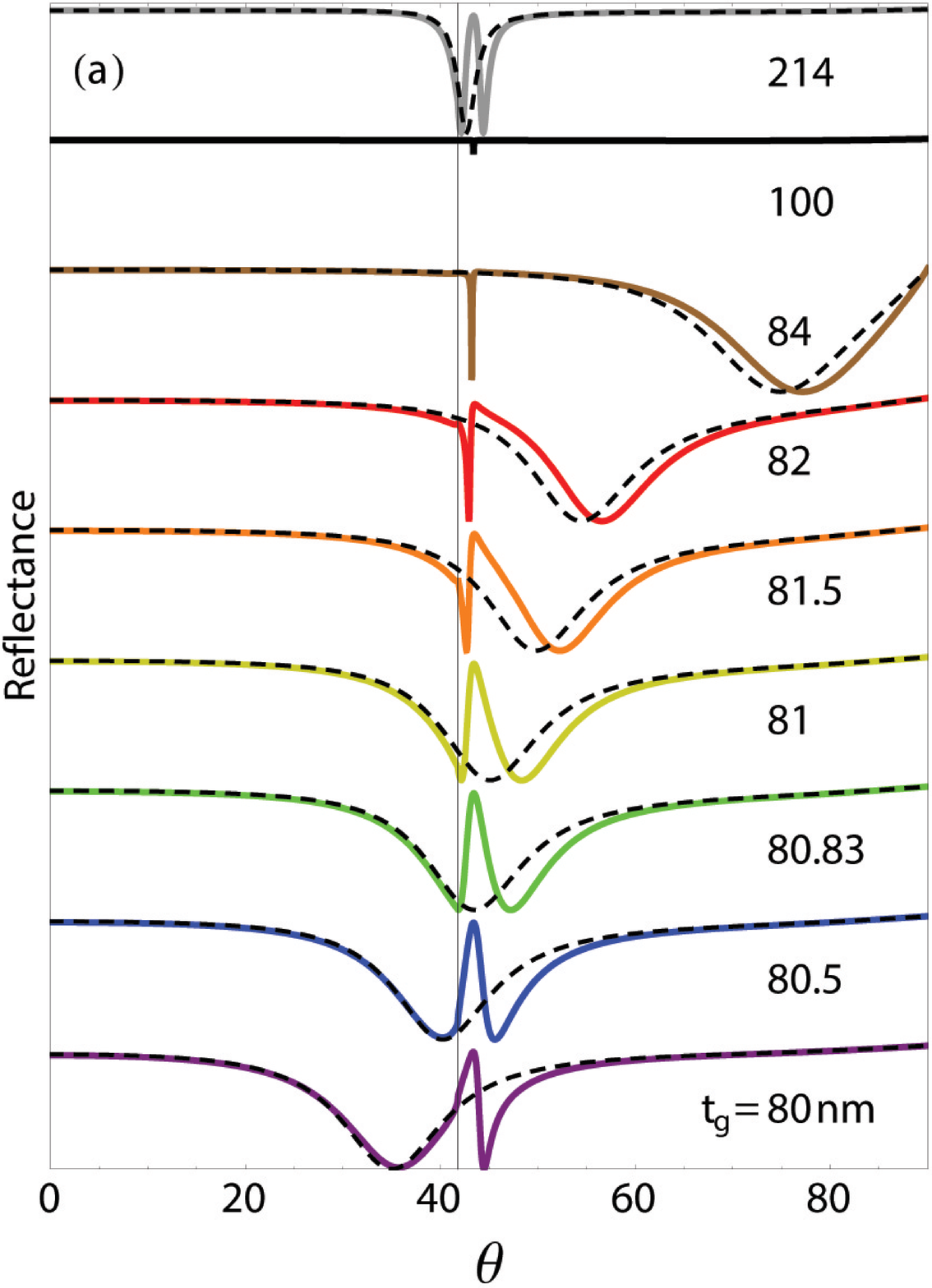}}
\end{minipage}
\begin{minipage}{.3\textwidth}
  \centerline{\includegraphics[width=\columnwidth]{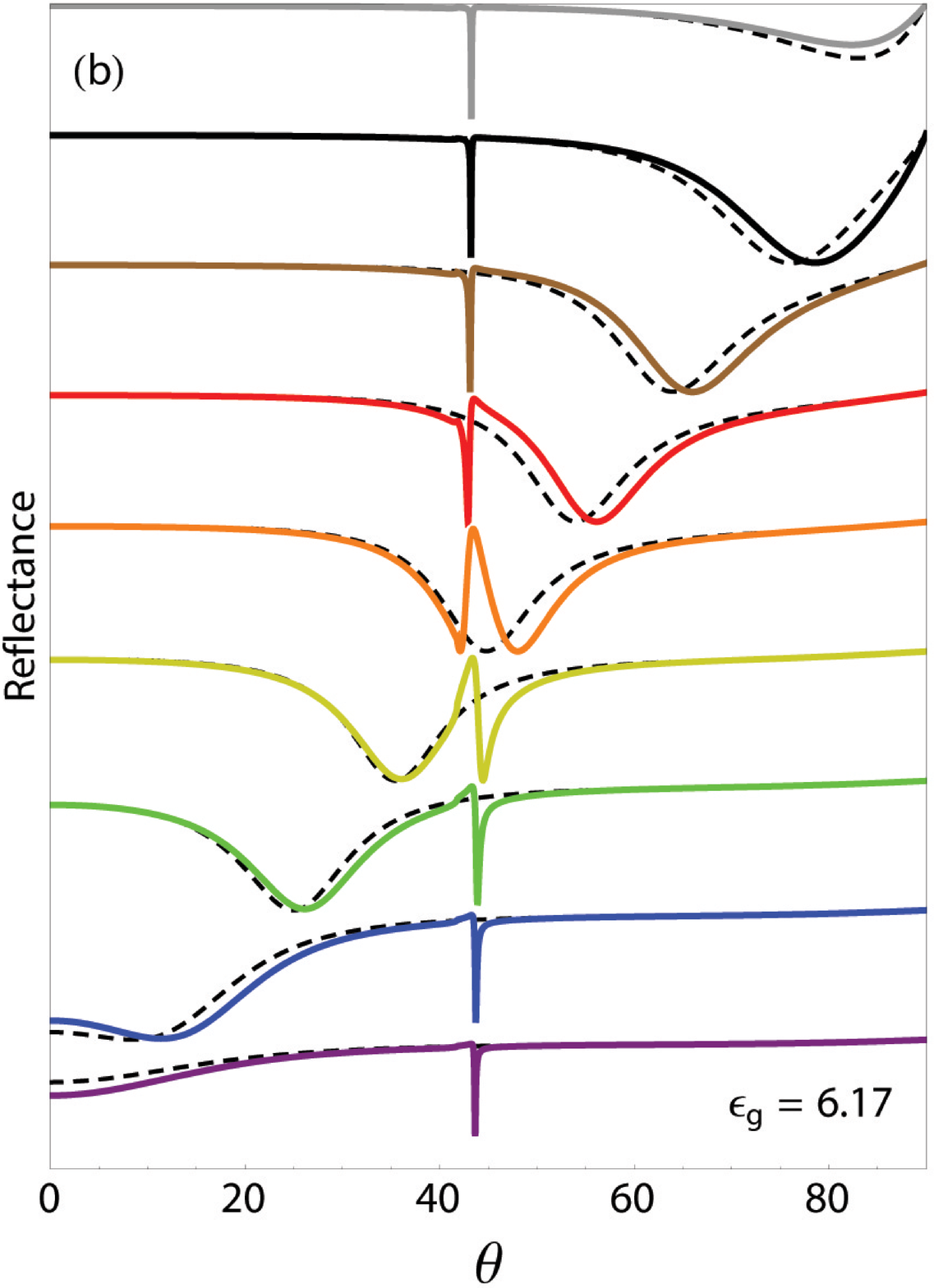}}
\end{minipage}
\begin{minipage}{.3\textwidth}
  \centerline{\includegraphics[width=\columnwidth]{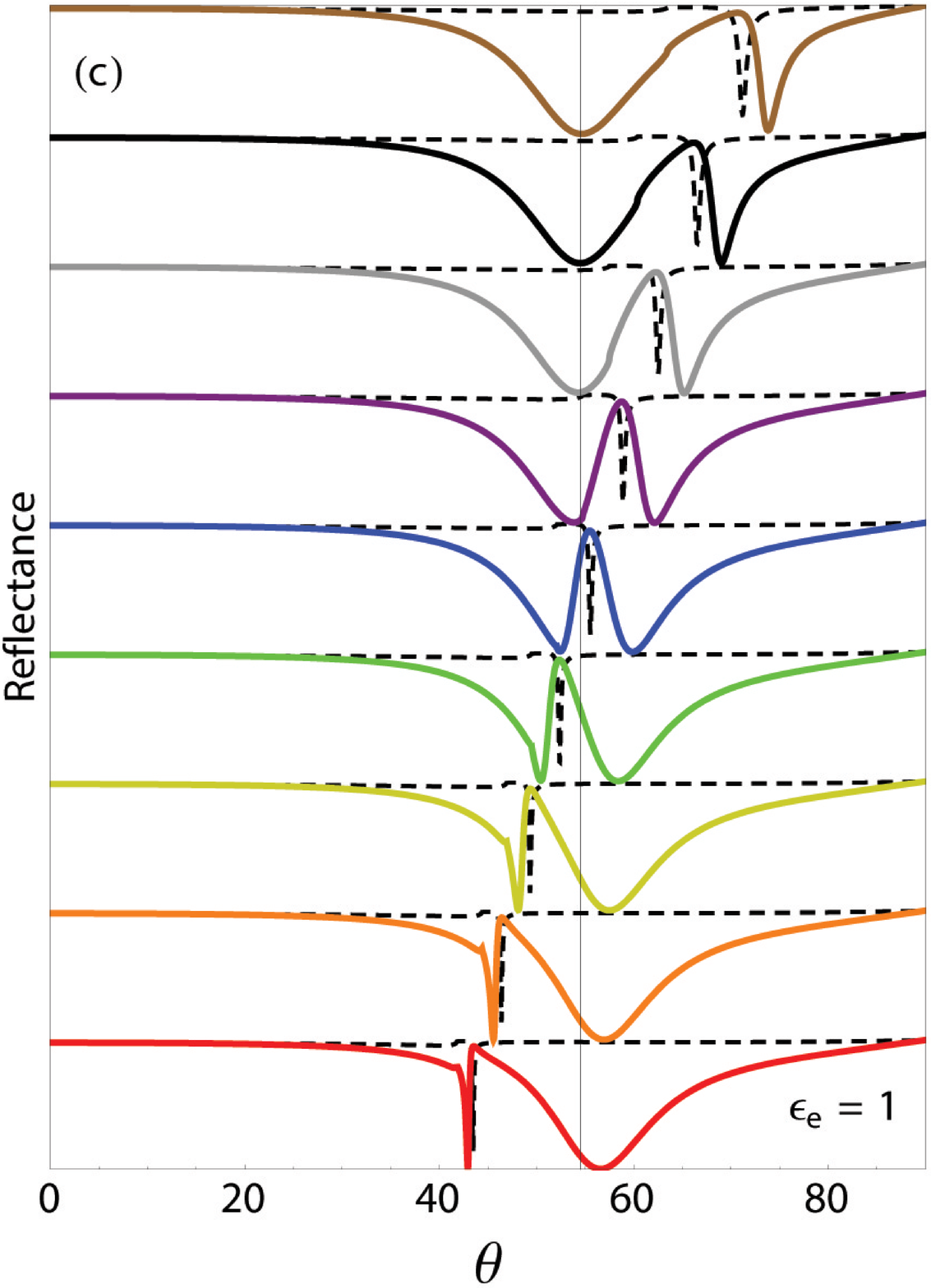}}
\end{minipage}
  \caption{Tuning of the GFR through variations of structural and materials parameters. Traces have been displaced vertically for clarity. All traces are shown on same scale. When tuning a permittivity, increments of 0.1 in value are used, with the lowest at the bottom of the panel. (a) Computed reflectance of MIS, showing tuning of the resonance through variation of the insulator thickness, as denoted. For this panel we use $\epsilon_g=\epsilon_{TiO_2}$ and $\epsilon_e=1$. All film thicknesses are in nm. The dashed curve underlying each trace designates the reflectance of semi-infinite MIM at the chosen film thickness. The critical angle is designated by a thin vertical line intersecting at $\theta=41.8^\circ$. (b) Computed reflectance of MIS, showing tuning of the resonance through variation of $\epsilon_g$ while using $t_g=80$nm and $\epsilon_e=1$. The dashed traces show the semi-infinite MIM reflectance. (c) Computed reflectance of MIS, showing tuning of the resonance through variation of $\epsilon_e$. The insulator thickness is set at $t_g=82$nm, and $\epsilon_g=\epsilon_{TiO_2}$. The dashed traces indicate the SPR for each permittivity value. The thin vertical line indicates the resonance angle of the semi-infinite MIM.}
  \label{fig:GFRtune}
\end{figure*}

We now address tuning of the GFR through geometrical and materials parameters. As discussed above, increasing the optical thickness of the insulator film redshifts the MIM mode, while $\theta_{cr}$ and the SPR remain unperturbed. This enables tuning of the GFR across the entire visible frequency range, as seen in Fig.~\ref{fig:modes}(c). The redshift is also accompanied by an increase in the MIM wavevector eignevalues at the fixed operating frequency of $\omega/\omega_p=0.216$.  We thus find that increasing the optical thickness distorts and spreads the GFR over an increasingly broad range of angles. This is demonstrated in panels (a) and (b) of Fig.~\ref{fig:GFRtune}. For comparison we also plot the MIM resonance curve for each parameter value, indicated by the dashed traces in these panels. For thick enough insulator films, as exemplified by the trace at $t_g=100$nm in ~\ref{fig:GFRtune}(a), there exist no MIM eigenvalues accessible in the Kretschmann configuration. In this case the MIM mode is rendered dark, and only a very weak and strongly screened SPR is excited. As the insulator thickness is further increased, higher order MIM modes may be excited. One such mode, along with its associated GFR is shown for $t_g=214$nm in Fig.~\ref{fig:GFRtune}(a). These higher order resonances are narrower than their fundamental counterparts, as expected.

The last control parameter we examine is $\epsilon_e$. In Fig.~\ref{fig:GFRtune}(c) we plot a series of GFR traces obtained using a range of values for the permittivity of the external insulator. The dashed trace underlying each GFR shows the SPR for each permittivity value. We find that when $\epsilon_e$ is varied, as would often occur in a sensing configuration, the GFR again undergoes angular tuning. In particular, increasing $\epsilon_e$ shifts the SPR to higher angles, in the process sweeping it across the now weakly perturbed MIM resonance.

An interesting feature that emerges while tuning the GFR is the reversal of its asymmetry, which occurs as the resonance is tuned across a range of angles. Close examination of the resonance curves in Fig.~\ref{fig:GFRtune} reveals that this takes place when one of the two interfering resonances is swept across the other, by varying either the optical path length or the external permittivity. This effect is known as \emph{q}-reversal, where \emph{q} is the Fano lineshape parameter~\cite{Connerade1985}. Such symmetry reversals have been observed in both quantum and classical interfering systems~\cite{Fan2002, Kobayashi2002}, and are generally attributed to the presence of an additional interfering channel, known as an \emph{interloper}, which is responsible for the additional phase variation~\cite{Yoshihara1993, Sadeghpour2007}. Due to the challenges associated with creating isolated, two-channel interfering systems, interlopers are often present in Fano systems. In the case studied here, we believe the interloper to be the continuum channel a-priori interfering with the surface-bound wave, to create the SPR mode. This channel consists of a broad and flat continuum, comprising radiation modes that undergo total internal reflection. Quantum interference of this radiation field with the surface mode field results in the characteristic SPR angular lineshape~\cite{Herminghaus1994}. Thus, the composite nature of the SPR naturally provides an additional interloper channel, explaining the observed \emph{q} reversal.

In conclusion, we have demonstrated the emergence of a geometric Fano resonance in a metal/insulator stack. Tuning of the resonance, either in the spectral or angular domain is achieved through control of geometrical and materials parameters. In vicinity of the SPR angle, momentum conservation of the reflected fields renders the excitation of the SPR and MIM modes indistinguishable, resulting in an angular coherence. When the SPR and MIM excitation angles are equal, the angular reflectance exhibits an EIT-like peak. Variations in the tuning parameters also results in \emph{q} reversal. We attribute this phenomenon to the presence of an interloper -- a broad continuum channel that underlies the SPR mode.

This work was supported by NSF grants ECCS-1202182 and DMR-1404676. H. Grotewohl acknowledges support of HHMI Science Education Award 52006956.

\end{document}